\documentclass[preprint2,10pt]{aastex}

\shorttitle{Chandra View of Abell 3266}
\shortauthors{Henriksen and Tittley}

\begin{document}

\title{Chandra Observations of the A3266 Galaxy Cluster Merger}

\author{Mark J. Henriksen\altaffilmark{1} and Eric R. Tittley}
\affil{Joint Center for Astrophysics, Physics Department, University of Maryland, Baltimore County,
    Baltimore, MD 21250}

\altaffiltext{1}{Also Laboratory for High Energy Astrophysics, NASA/GSFC} 

\begin{abstract}
Analysis of a 30,000 s X-ray observation of the Abell 3266 galaxy cluster with 
the ACIS on board the {\it Chandra} Observatory has produced  several new
insights into the cluster merger. The intracluster medium has a
non-monotonically decreasing radial abundance profile. We argue that the most
plausible origin for the abundance enhancement is unmixed, high abundance
subcluster gas from the merger. The enrichment consists of two stages:
off-center deposition of a higher abundance material during a subcluster merger
followed by a strong, localized intracluster wind that acts to drive out the
light elements, producing the observed abundance enhancement. The wind is
needed to account for both an increase in the heavy element abundance and the
lack of an enhancement in the gas density. Dynamical evidence for the wind
includes: (1) a large scale, low surface brightness feature perpendicular to
the merger axis that appears to be an asymmetric pattern of gas flow to the
northwest, away from the center of the main cluster, (2) compressed gas in the
opposite direction (toward the cluster center), and (3), the hottest regions
visible in the temperature map coincide with the proposed merger geometry and
the resultant gas flow.  The Chandra data for the central region of the main
cluster shows a slightly cooler, filamentary region that is centered on the
central cD galaxy and is aligned with the merger axis directly linking the
dynamical  state of the cD to the merger.  Overall, the high spectral/spatial
resolution Chandra observations  support our earlier hypothesis (Henriksen,
Donnelly, \& Davis 1999) that we are viewing a  minor merger in the plane of
the sky.
\end{abstract}


\section{Introduction}
Galaxy clusters show a wide range of complex phenomenon in the X-ray regime including 
radial abundance gradients, 
radial temperature gradients, cooling flows, two-phase gas in the central region (Makishima et al. 2001), 
non-thermal emission, and two dimension features such as shock fronts.

Observations with {\it ROSAT}, {\it ASCA}, and {\it Chandra}
have shown that cluster evolution proceeds through mergers and that complex physical processes 
account for the presence or absence of many of these observed phenomenon. With the 
data sets now becoming available, we are in a position to 
address the physics of cluster evolution. {\it ASCA} observations
of cluster mergers, such as Abell 754 (Henriksen \& Markevitch 1996) and Coma 
(Watanabe et al. 1999), offer a coarse 
description of the pattern of shocked gas that generally matches the merger scenarios 
played out in N-body hydrodynamical simulations.
However, with increased resolution, {\it Newton-XMM} and {\it Chandra} have begun to provide
results that go beyond the current simulations of cluster evolution and may
perhaps require additional physical processes. For example, the distribution
of metals in the intracluster medium (ICM)
is an important diagnostic for galaxy evolution within clusters and for the cluster gas dynamics. 
Two-dimensional abundance maps are within the capabilities of both {\it Newton-XMM} and {\it Chandra},
yet the distribution of metals in the intracluster
medium has not been simulated.

Abell 3266 has become a well studied cluster merger first shown to have evidence for a merger 
(in addition to optical substructure) based on detection of 
shocked gas through detailed modeling of {\it ASCA} observations (Henriksen,
Donnelly, \& Davis 1999). Mergers are not always apparent in the
optical data and clusters may even appear to have a 
``relaxed" morphology, either spatially or dynamically. 
For example, though A3266 has spatial substructure, it
looks dynamically ``quiet" in the optical in that it
has a velocity distribution that is statistically consistent with a single Gaussian.  
On the other hand, the {\it ASCA} temperature
map and the {\it Chandra} temperature map unambiguously show high temperature regions
that may be associated with shocked gas. 
With a detailed statistical analysis 
of the galaxy positions and redshifts obtained by Quintana et al. (1996), 
in an earlier paper we found that the cluster is composed of a secondary 
component comprised of approximately 30 galaxies ($\sim$10\% of the
total number of galaxies with measured redshifts). The velocity dispersion
is significantly higher, $\sim$1300 km s$^{-1}$ in the center 
compared to the global value of
1000 km s$^{-1}$. This was interpreted as evidence 
of a merger in progress.
Further evidence of a merger comes from the co-alignment of the  
primary and secondary galaxy condensations with the elongation
in the X-ray
contours first noted by Mohr, Fabricant, \& Geller, (1993). The elongation
extends from the arc minute scale in the {\it ROSAT} PSPC image to the location of the secondary 
galaxy condensation, 15 $\arcmin$ to the NE, and is
visible in the {\it ASCA} GIS image. This, together with the Gaussian velocity distribution, 
suggests a merger in the plane of 
the sky along this axis of elongation.
The temperature discontinuity in the {\it ASCA} temperature
map
is perpendicular to the 
axis connecting the subclusters suggesting that the subcluster to the NE has
already traversed the core and partially shockheated the ICM. Application of the 
pre- and post-shock gas temperatures
derived from the {\it ASCA} temperature map to the shock equations give a post-shock gas velocity of
1400 km sec$^{-1}$. Simulations of off center cluster mergers occurring
into the plane of the sky (Takizawa 2000), 
show a similar temperature morphology
to the {\it ASCA} map of A3266 but require a larger secondary
component relative to the primary (1:4) than results from our galaxy component
separation ($\sim$1:10). Thus, the determination
of merger geometry is somewhat ambiguous using only the 2-dimensional temperature map. 
However, a secondary component this large falling into the plane of the sky should appear as a 
spike in the velocity distribution and significantly distort the Gaussianity of the velocity
dispersion.
This is not seen in our statistical analysis that shows the inner region, outer region,
and entire cluster are all consistent with a single Gaussian profile with a similar heliocentric
velocity peak (Henriksen, Donnelly, \& Davis 1999).
Additionally, there is a narrow angle tail (NAT) and a
wide angle tail (WAT) radio source with velocities well 
within the cluster velocity dispersion and located, in 2-dimensions, within the
shocked gas. They are oriented with their tail aligned with the direction of the gas flow. 
Therefore, a major merger into the plane of the sky seems less likely than a minor merger in the plane of the
sky.
This point is discussed in more detail in light of the new results from {\it Chandra} in Section 4.2

{\it BeppoSax} observations showed that the cluster, overall, exhibits a radially decreasing temperature 
profile (De Grandi \&
Molendi, 1999). {\it Chandra} 
provides a more detailed description of the central region
and gives a temperature map that is not
subject to the PSF deconvolution techniques used with {\it ASCA} data (Section 3.1).
The {\it BeppoSAX} data did not show any departures from constant metallicity
out to 10$\arcmin$. Using Chandra, we are able to resolve the abundance 
within the central resolution element of the BeppoSax observations, $\sim$2 arcmin (Section 4.1),
and uncover an enhancement. 
These results support the merger hypothesis in Abell 3266 and
give a more detailed picture of the dynamics of the ICM during a merger. Finally, the cD 
galaxy in Abell 3266 has multiple nuclei. Multiple nuclei are found in
28\% of the first rank galaxies in rich clusters (Hoessel 1980). The {\it Chandra} observations 
provide unique evidence that 
the formation of
the multiple nucleus cD in Abell 3266 
may be linked to the merger (Section 4.3). All parameters are quoted with 90\% confidence
errors. A Hubble constant of 60 km s$^{-1}$ Mpc$^{-1}$ is used throughout.

\section{Observations and Analysis}
Abell 3266 (also known as Sersic 40-6), located at  (J2000) 04h30m22.0s, -61d32m03s with redshift of 
0.0594, was observed on 
July 26, 2000 for 30,000 seconds. 
The cluster is centered in chip 1 of the ACIS-I detector and the emission spans all four
chips. The total count rate is 10.56 $\pm$0.02 cts s$^{-1}$, well below the telemetry saturation limit. 
Figure 1 is a color coded 
intensity map that shows the full 22$\arcmin$ square image in the 0.3 - 8 keV band.
The image was obtained using processing routines
outlined in the CIAO 2.1 Science Threads. The events data were first cleaned for flares.
The intensity map was then background subtracted and
divided by the exposure map. 
The background map was created following the CIAO 2.1 Science Thread, 
``ACIS Blank-Sky Files" and uses as its basis a blank-sky dataset compiled by Maxim Markevitch.
The background map and intensity map were normalized by the exposure times of the data sets.
The normalization was verified by 
comparing the high-energy tail of the background events with the tail in the distribution for the study data,
which should have little emission at the highest energies.

This intensity map is quite complex and shows several interesting features.
Overall, the cluster is similarly elongated as in the {\it ROSAT} and {\it ASCA} GIS
images in the direction toward the secondary optical subcondensation to the NE. 
To the NW, there is a steep drop in intensity over 2 - 3$\arcmin$. The
intensity peak is visibly displaced to the NW. The {\it ROSAT} PSPC emission peak was within 
30 arcsec of the cD galaxy but the {\it ROSAT} pointing error did not allow a clear identification. 
Figure 2 shows the X-ray contour
map from the ACIS observation overlaid on the DSS image. The contours are
1, 1.5, 2, 2.5, 3, 3.5, and 4$\times$10$^{-4}$ cts cm$^{-2}$ s$^{-1}$ arcmin$^{-2}$. It is apparent 
from this figure that
the X-ray peak is co-spatial with the cD galaxy, (J2000) 04h31m12.16s, -61d27m15.3s. 

Spectra were extracted from  the events file in regions of interest.  The spectra are grouped so that there are no fewer than 64 events per grouped bin.
A weighted spectral redistribution matrix (RMF) and effective area function (ARF) for each region was found using the {\tt calcrmf} and {\tt calcarf} 
tools, version 1.04. The mean RMF (and ARF) is computed from the weighted average of the RMFs (ARFs) at the different positions in the region.  The weights are derived from a coarse-grained image of the data in the region of interest.  

\section{Results}
A thermal model was fit to the spectrum extracted from a circular region of 2.5$\arcmin$ centered at
the cluster center, (J2000) 04h31m21.04s, -61d27m00.16s, for comparison to previous observations. The
results are given in Table 1. The temperature range, 9.17 - 9.60 
keV, is consistent with the central region as
observed by {\it BeppoSax} (De Grandi \& Molendi 1999). The {\it Chandra}
 central abundance, 0.19 - 0.33 (90\% 
confidence), is consistent with the BeppoSax
value, 0.22 - 0.34 (68\% confidence). The emission weighted 
temperature obtained from joint fitting the {\it ROSAT} PSPC and {\it ASCA} GIS observations 
taken from a circular
region of radius $\sim$17 $\arcmin$ is 7.75 - 8.46 keV while the abundance is 0.15 - 0.25 Solar
(Henriksen, Donnelly, \& Davis 1999). The combination of 
PSPC and GIS gives a similar energy
band to the ACIS, similar sensitivity above 2.5 keV, and significantly higher sensitivity below 2.5 keV. The lower 
temperature of the large region compared to the small central region is indicative of non-isothermallity; in
particular, it requires cooler gas outside of the central region.
Radial temperature (see Figure 3) and abundance (see Figure 4) profiles were made centered on the ACIS 
X-ray peak. 
The abundance and temperature measurements corresponding to the annular regions
are given in Table 2. Cooler gas at large radii is detected as expected from the earlier comparisons with the PSPC/GIS spectra.
The temperature decreases steeply at $>$7$\arcmin$ as opposed to smoothly decreasing 
as would be expected
from a gas in hydrostatic equilibrium. The steep temperature ridge may coincide with the edge of the shock front.
The {\it Chandra} radial temperature profile in the central region
is in general agreement with the {\it BeppoSax}
observation, however, the ACIS analysis allows smaller radial bins and 
shows a shallow increase in temperature within
the central 3$\arcmin$. The central 1$\arcmin$ radius circle is centered on the brightness peak and likely 
has 
a contribution from the central cD. Cooler gas, $\sim$ 1 - 2 keV, typical of a cD 
galaxy and the hot intracluster gas likely account for the lower
observed temperature.

The radial abundance map shown in Figure 4 spans a radius of $\sim$10$\arcmin$. 
The best fit Sulphur abundance (1.0 +/- 0.7 Solar) is consistently $\sim$3 times higher than
the average abundance of the ICM (0.26 +/-0.07 Solar) though they are consistent
within the large error of the Sulphur abundance. The Sulphur-to-Iron
ratio is predicted to be quite high for SNe II enrichment compared
to that of type SNe Ia (along with Si, Ne, O).
However, it appears to occur at energies just above the
Iridium edge of the Chandra
mirror around 2.2 keV where the effective area shows a sharp
discontinuity in the ACIS response matrix that may depress 
the continuum near the redshifted sulphur line raising the inferred Sulphur abundance.
Therefore, the metallicity
was measured tying all of the elements except Sulphur
which is left as parameter. The radial abundance measurements are inconsistent with constant
abundance. There is an overall constant abundance with the second annulus from the 
center enhanced at double the 
cluster average. The possibility that the abundance in the inner region
is low due to multi-phase gas that is poorly modeled with a single component as was found for
M87 ( Molendi \& Gastaldello 2001) was investigated. Refitting the spectrum with a second temperature component
does not improve the fit. The spectrum and residuals from fitting this region with a single
temperature component are shown in Figure 5. If a significant cool temperature component
were present and not properly modeled, there should be significant positive residuals around FeL
from the cool component. This is not seen in the spectrum.
The emission was also modeled in 4 quadrants to look for asymmetries in the abundance, 
temperature, and emission integral (Figure 6). These results are given in
Table 3. The temperature does not exhibit azimuthal
variations. The abundance has large error bars, 30 - 50\%; variations of
$>$50\% are not seen. Only the emission integral shows variations 
of 50\% due to the visibly asymmetric gas distribution. The column density is left as
a free parameter and is significantly higher
than the measured galactic column density, 1.6$\times$10$^{20}$ (Dickey \& Lockman 1990).
We refit the spectrum excluding data below 1.5 keV to minimize the effect of
galactic absorption on the spectrum. The best fit temperature and abundance are consistent
with those measured using the data above 0.5 keV. The high measured galactic column
densities do not seem to bias the values obtained for the temperature and the abundance.
We also found, by fitting regions spanning the entire ACIS-I array, that the high fit n$_{H}$
is not local to the central region of the cluster but spans the entire cluster region or CCD
array. Thus, the high n$_{H}$ values may be due to a poor calibration below 1.5 keV for this
observation.

\subsection{2-Dimensional Temperature Map}
The ACIS-I observation spans a large part of the central and western side of the cluster as shown in the temperature
map with X-ray isointensity contours overlaid (see Figure 7). The regions are numbered with the corresponding
temperatures given in Table 4. The temperature map is color coded to facilitate recognition of broad trends.

To determine the regions to be analyzed, the events data for the chips 0-3 were 
processed through a simple algorithm which generates 
a set of regions with a quasi-uniform number of events per region.  For a given number of events, 
$N$, the algorithm tests if more than $4N$ events occur in a region. If there are more than this 
number, then the region is split into 4 equal-area regions. The regions in Figure 7
were created using $N=4096$.  Only four of the 28 regions created had less 
than 4096 events (12, 11, 15, and 7) with region 12 having the least with 2632 events.  Nine regions have 
4000 to 6000 events, and the remaining 15 regions are evenly spread among having 6000 to 15000 events.

A qualitative comparison with the 
{\it ASCA} temperature map in Henriksen, Donnelly, \& Davis (1999) can be done since regions 
5,6, and 8 on the {\it ASCA} map are
resolved into smaller regions in the {\it ACIS} map. The hottest region in the {\it ASCA} map, 
11.0 - 16.7 keV, is region 8 which is a small 
section to the southwest of the X-ray peak. The {\it ACIS} map
shows that the primary contributions to {\it ASCA} region 8 are two regions, with temperatures at 11 and 12 keV, 
color coded pale 
yellow and white, respectively.
Region 6 in the {\it ASCA} map is 9 - 10.5 keV and is located to the southeast of the X-ray peak. 
While the {\it ASCA} map extends several arc minutes below
the overlapping region on the ACIS, the ACIS shows 4 contributing regions with 
temperatures of 7, 8, 10 and 11, keV. The average
of these 4 regions, 9 keV, is consistent the temperature of the corresponding {\it ASCA} region. Region 5 on the
{\it ASCA} map is slightly cooler, 8.6 - 9.6 and extends to the northwest of the X-ray peak. 
While the ACIS map shows a 
couple  of small hot regions with temperatures of 11 and 12 keV, however,
the dominant contribution is from a large region with an average temperature of 8 keV. Interestingly,
region 3, cooler regions in the {\it ASCA} temperature map, 7.4 - 8.5 keV, is slightly east of the center 
and extends to the northwest.
The ACIS map shows that this region is dominated by two regions at 7.3 and 8.5 keV.
Thus, it is apparent that the 
temperature map derived from the ACIS, with its better spatial resolution, is consistent with the temperature
map derived from {\it ASCA}. The geometry of the shock front is similar in 
both the ACIS and {\it ASCA} maps and generally runs from 
north to south and is centered to the west of the
X-ray peak. 

\section{Discussion}

\subsection{Abundance Gradients
 and Gas Mixing}
Cluster
abundance gradients, including Perseus (Ezawa et al., 2001), S\'ersic 159-03 (Kaastra et al. 2001), A496 
(Dupke \& White, 2000), AWM 7 (Ezawa et al. 1997), and the mean 
metallicity profile for cooling flow clusters (De Grandi \& Molendi, 2001), 
are usually described as monotonically decreasing radial profiles. 
Abundance gradients may preferentially occur in clusters with cooling flows (Allen \& Fabian 1998)
and the BeppoSax survey of De Grandi \& Molendi tends to support this idea.
Cooling flows are
generally thought to be disrupted by mergers.
Gas mixing due to a merger would erase the gradient as it disrupts the cooling flow
so that abundance gradients
and cooling flows should anti-correlate with a merger. 

However,
there are several clusters with a non-constant abundance (reported 
in the {\it BeppoSax} survey) without
cooling flows: Abell 119, Abell 3627, and the Coma cluster. Abell 119
has a very similar radial abundance profile to Abell 3266 (as reported here)
while Abell 3627 and Coma
appear to have an abundance deficit in the center of an otherwise flat profile.
The abundance change seen in Figure 4 shows that there is a significant ($>$90\%
confidence) difference between the second annulus and the third, fourth, and fifth annuli.
This appears to be an enhancement to an otherwise flat abundance profile.
In the case of Abell 3266,
the 
subcluster appears to have deposited sufficient 
metals to raise the observed abundance
to 0.4 Solar in the second annulus. Dumping these metals would also increase the gas density
in this annulus. We used the archival 
ROSAT PSPC observation of 13,288 seconds to obtain a radial profile and performed a
beta-model fit to it to look for the density enhancement. The beta-model
is given by S(r) = S$_{c}$(1 + (r/a)$^{2}$)$^{(-3\beta + 1/2)}$. The surface brightness
is described by S with the central value given by S$_{c}$. The core radius (a)
and $\beta$ are the other free parameters in the fit. Our temperature
map and radial profile indicate that the cluster gas is dominated by gas at $>$6 keV. The PSPC surface brightness should be insensitive
to temperature changes in the gas and accurately map the density changes.
The image was flatfielded using the exposure map
and background was subtracted before extracting the radial profile. 
Figure 8 shows a beta model fit out to a radius of 10$\arcmin$ ($\sim$ 1 Mpc)
from the cluster center.
The beta model provides a statistically acceptable fit, $\chi^{2}$ $\sim$1, 
to the data within this region,
however, fitting data at larger radius gives a very poor fit due to the pronounced asymmetry
in the surface brightness that increases with radius. Confidence contours
for beta and the core radius are shown in Figure 9. The range of beta and core radius
are typical of clusters observed with ROSAT (Vikhlinin et al. 1999). In the
annular region where the abundance enhancement is observed 
with the ACIS, r = 0.9 - 1.8 arc min, no significant surface brightness enhancement is seen
above the beta-model. This indicates that there is no significant increase in gas density
relative to the surrounding annular regions. An increase in abundance without a change in
gas density requires that metals are dumped in and retained while the bulk
of the gas is transported away.  

We suggest that the 
energy input from shock heating would preferentially transport the light 
ions (e.g., H and He) out of the subcluster penetration region since the
heavy ions are much more massive and therefore have a much lower velocity.
This increases the
metal abundance since the metal number density is measured relative to Hydrogen.
Figure 10 shows a drawing of the heating that is visible in simulations of
the cluster merger using the Hydra code (Couchman, Thomas, \& Pearce 1995) that we
found matches our proposed merger geometry (in Section 4.2). As the subcluster penetrates 
the ICM of the main cluster along the proposed merger path,
it drives shocked gas toward the cluster center and toward the NW away from the
center. In the post-shock gas,
light ions and electrons will diffuse out more quickly  
accounting for the preferential retention of heavy ions in the region of subcluster 
gas deposition. In the presence of a weak, chaotic cluster magnetic field, thermal conduction will
still be efficient (Narayan \& Medvedev, 2001) and this formation process for
the abundance enhancement remains plausible.
Additional support for this scenario is apparent in the temperature map
that shows several of the hottest 
regions located 1 - 2$\arcmin$ from the cluster center at the approximate location of the shocked gas.
Figure 2 also shows the compression of gas along the
NW face of the central region as the shock front moves toward the SE. Figure 1 which shows
``fingers of X-ray emission" reaching to the NW, presumably expanding post-shock gas.

\subsection{Merger Geometry and Viewing}
The {\it ACIS} temperature map shows much more detail than was visible in the {\it ASCA} temperature
map. Figure 7 shows the X-ray surface brightness contours overlayed
on the color-coded temperature map which provides a spatial  comparison
of the relative gas density to gas temperature. Five resolved regions with both high temperature,
$\ge$ 10 keV, and high density generally lie to the 
north, west, and south of the X-ray peak. The highest temperature regions, $\ge$11.5 keV, coincide with a 
steep density gradient that traces out the high entropy, post shock gas and generally runs north to south.
In contrast, the cluster peak and the region to the northeast is cooler, $\sim$8 keV, and has a lower density. 
These general characteristics are crucial for determining the merger geometry.
Roetigger and Flores (2000) make predictions for the temperature map assuming a merger 3 Gyr 
ago, off axis, at
an angle of 45$^{o}$ to the plane of the sky,
with a mass ratio of 2.5:1. While their predicted temperature map (their Figure 3) 
is qualitatively consistent with the ASCA temperature map, a significant difference becomes
apparent with the better spatial resolution of Chandra. Their model predicts that the X-ray peak should coincide with the 
highest temperature gas. The ACIS map shows this is not the case. We propose our Figure
10 as the merger geometry that is consistent with the temperature map. 
Another problem with the proposed merger model of Roetigger and Flores 
(in which a subcluster is falling into the plane of the sky at 45 degrees with
the mass ratio of 2.5:1) is that there should be a significant asymmetry on the high velocity side
of the velocity dispersion. 
The proposed line-of-sight component
to the bulk infall velocity, $\sim$1800 km s$^{-1}$, far exceeds the velocity dispersion, 
$\sim$1000 km s$^{-1}$, of the entire cluster (Henriksen,
Donnelly, \& Davis 1999). Referring to our previously published
velocity histogram, in the region of 19,600 km s$^{-1}$, the heliocentric velocity of 
the primary cluster (17,804 km s$^{-1}$) plus
the predicted infall velocity of secondary component (1800 km s$^{-1}$), 
there are only $\sim$30 galaxies (in velocity space) or 
$<$10\% of the galaxies.
This would seem inconsistent with a 2.5:1 mass ratio for the proposed merger that 
would predict closer 
to 100 galaxies in the infalling component.
Finally, the {\it ACIS} map shows cool gas running along the merger 
axis (see Figure 10). The relationship of this gas to the X-ray peak and the cD galaxy, 
discussed in more detail below, is most naturally
explained by a merger in the plane of the sky.

\subsection{Formation of Double Nucleus cD}
The X-ray
intensity peak is slightly displaced to the NW of the visual centroid of the X-ray emission. 
The X-ray peak is co-spatial with the cD galaxy, (J2000) 04h31m12.16s, -61d27m15.3s. 
The cD has a secondary nucleus with a relative velocity
of $\sim$400 +/- 39 km$^{-1}$. This is much larger than the stellar velocity dispersion of the cD nucleus, 
$\sim$327 +/- 34 km$^{-1}$ . In addition, there
is an asymmetric rise in the stellar velocity dispersion of the cD that peaks at 700 km$^{-1}$ (Carter et al. (1985)
suggesting that the cD is tidally disturbed by a very massive object, perhaps the cluster merger. 
Heating in the stellar velocity dispersion is greater than the relative velocity of the nuclei themselves which
makes it unlikely that the double nucleus interaction is responsible.
The {\it Chandra} data
build the case for the scenario that the formation of the
dumbbell morphology is likely due to the cluster merger. Figure 11 is an X-ray contour overlay on the 
color-coded hardness 
ratio of the central 6 x 12.5$\arcmin$ region.
The coolest region runs along the axis of the surface brightness elongation running SW to NE. This $\sim$2.5$\arcmin$ plume has a nonlinear
shape with its apex at the cD galaxy location. Alignment of the plume with the merger axis
provides a link between the cD galaxy and the merger. Tidal heating in the cD and the relative velocity of the
nuclei are so large that they can only be attributed to the merger. This plume of cooler gas could to be stripped material from the cD galaxy,
or a disrupted
cooling flow centered on the galaxies. 
The second nucleus may have been captured from
the subcluster on its initial pass accounting for the pre-equilibrium, high relative velocity of the two nuclei. 
The standard picture for the formation of a double nucleus is that it is formed by cannibalizing neighboring galaxies. In
the case of A3266, the preponderance of data (e.g., the high relative velocity, extreme stellar heating, and most importantly
the plume that links the cD to the merger) seem to indicate cannibalizing galaxies from a merging subcluster.

\section{Conclusions}
The {\it Chandra} observations of Abell 3266 provide a detailed picture of the cluster merger. Evidence was given to support
the hypothesis that the merger is in the plane of the sky and is a relatively minor one. These observations
show that the merger is undoubtedly linked to the dynamical state of the dumbbell galaxy morphology. An off center
abundance enhancement was found that we suggest was formed by the merger. The formation process
consists of deposition of higher metallicity material during the merger
followed by a cluster wind that reduces the gas density while preferentially retaining the metals. Future observation with ASTRO-E2 XRS 
will be especially interesting for Abell 3266 since they will provide gas velocities
that will directly test our conclusions about the merger geometry and formation of the abundance profile.

\acknowledgments
This work was supported by SAO Grant No. GO0-1154X.

\begin{deluxetable}{crrrr}
\tablecaption{Single Region Temperature Fit. \label{tbl-1}}
\tablewidth{0pt}
\tablehead{
 \colhead{kT\tablenotemark{a} (keV)}          & 
 \colhead{Abundance\tablenotemark{b}}   &
 \colhead{S Abundance\tablenotemark{c}} &
 \colhead{nH ($10^{20}$cm${^{-2}}$})&
 \colhead{Normalization}
} 
\startdata
8.6 $\pm$0.5 & 0.26 $\pm$0.07  & 1.0 $\pm$0.7& 9.9$\pm$0.8 & 1.22 $\pm$ 0.02 \\
\enddata
\tablenotetext{a}{Emission weighted temperature}
\tablenotetext{b}{Abundance (excluding S) relative to solar}
\tablenotetext{c}{Sulphur abundance relative to Hydrogen: $1.62\times 10^{-5}$,  Anders E. \& Grevesse N. (1989)}
\tablenotetext{d}{$10^{-14} \over 4\pi ((D(1 + z))^2$ $\int$ n$_e$n$_H$dV in cm$^{-5}$}
\end{deluxetable}

\begin{deluxetable}{crrrrr}
\tablecaption{Parameters for Annular Regions. \label{tbl-2}}
\tablewidth{0pt}
\tablehead{
\colhead{Radius} & \colhead{Normalization\tablenotemark{a}}   & \colhead{Area\tablenotemark{b}}   
& \colhead{kT\tablenotemark{c}} & 
\colhead{Abundance\tablenotemark{d}} 
} 
\startdata
$<0.9\arcmin$           & 0.00231 $\pm$0.00009 &   2.74 & 8.90 $\pm$0.92 & 0.21 $\pm$0.19  \\
$0.9\arcmin-1.8\arcmin$ & 0.00420 $\pm$0.00007 &   7.92 & 9.45 $\pm$0.62 & 0.38 $\pm$0.14 \\
$1.8\arcmin-3.7\arcmin$ & 0.01030 $\pm$0.00012 &  31.79 & 10.0 $\pm$0.65 & 0.18 $\pm$0.09  \\
$3.7\arcmin-7.3\arcmin$ & 0.01162 $\pm$0.00014 &  69.64 & 9.56 $\pm$0.38 & 0.16 $\pm$0.09 \\
$>7.3\arcmin$           & 0.00765 $\pm$0.00013 & 159.67 & 7.09 $\pm$0.42 & 0.08 $\pm$0.10 \\
\enddata
\tablenotetext{a}{$10^{-14} \over 4\pi ((D(1 + z))^2$ $\int$ n$_e$n$_H$dV in cm$^{-5}$}
\tablenotetext{b}{arcmin$^{2}$}
\tablenotetext{c}{Emission weighted temperature in keV}
\tablenotetext{d}{Abundance (excluding S) relative to solar}
\end{deluxetable}

\begin{deluxetable}{crrrrr}
\tablecaption{Parameters for Azimuthal Regions. \label{tbl-3}}
\tablewidth{0pt}
\tablehead{
\colhead{Quadrant} & \colhead{Direction}   & \colhead{n$_H$\tablenotemark{a}}   & \colhead{kT} & 
\colhead{Abundance}  & \colhead{Normalization\tablenotemark{d}} 
} 
\startdata
1 & North & 9.7 $\pm$1.8 & 8.69 $\pm$0.76 & 0.30 $\pm$0.10 & 0.0124 $\pm$ 0.0004 \\
2 & South & 10.6 $\pm$2.3 & 9.13 $\pm$1.00 & 0.21 $\pm$0.13 & 0.0074 $\pm$ 0.0003 \\
3 & East &  10.4 $\pm$1.8 & 8.31 $\pm$0.76 & 0.29 $\pm$0.10 & 0.0108 $\pm$ 0.0004 \\
4 & West &  9.9 $\pm$2.4 & 9.42 $\pm$0.97 & 0.27 $\pm$0.13 & 0.0089 $\pm$ 0.0004 \\
\enddata
 \tablenotetext{a}{Column density of Hydrogen in 10$^{20}$ cm$^{-2}$}
\tablenotetext{d}{$10^{-14} \over 4\pi ((D(1 + z))^2$ $\int$ n$_e$n$_H$dV in cm$^{-5}$}
\end{deluxetable}

\begin{deluxetable}{crrrrr}
\tablecaption{Temperature and Abundance Map Data\label{tbl-4}}
\tablewidth{0pt}
\tablehead{
\colhead{Cell} & \colhead{kT (keV)}   & 
\colhead{Abundance}
}
\startdata
 1 &  8.7 $\pm$1.1 	& 0.00 $\pm$0.50 \\ 
 2 & 15.7 $\pm$9.9 	& 0.01 $\pm$1.23 \\ 
 3 &  7.2 $\pm$0.7	& 0.11 $\pm$0.31 \\ 
 4 &  8.1 $\pm$1.0  	& 0.24 $\pm$0.18 \\ 
 5 &  8.7 $\pm$2.0  	& 0.35 $\pm$0.15 \\ 
 6 &  7.3 $\pm$0.9 	& 0.35 $\pm$0.47 \\ 
 7 & 10.7 $\pm$2.8  	& 0.11 $\pm$0.50 \\ 
 8 &  8.9 $\pm$1.9  	& 0.01 $\pm$0.37 \\ 
 9 &  9.3 $\pm$3.0	& 0.19 $\pm$0.34 \\ 
10 & 10.8 $\pm$3.2 	& 0.37 $\pm$0.16 \\ 
11 & 15.0 $\pm$9.9  	& 0.08 $\pm$0.45 \\ 
12 & 11.2 $\pm$2.4 	& 0.35 $\pm$0.36 \\ 
13 & 10.3 $\pm$7.1  	& 0.67 $\pm$0.56 \\ 
14 &  9.0 $\pm$1.2	& 0.18 $\pm$0.35 \\ 
15 &  7.5 $\pm$1.4  	& 0.04 $\pm$0.18 \\ 
16 &  7.4 $\pm$2.0	& 0.00 $\pm$0.50 \\ 
17 &  8.5 $\pm$1.2 	& 0.23 $\pm$0.15 \\ 
18 &  7.2 $\pm$0.8 	& 0.14 $\pm$0.15 \\ 
19 &  6.5 $\pm$1.1	& 0.00 $\pm$0.50 \\ 
20 &  7.2 $\pm$1.2	& 0.10 $\pm$0.19 \\ 
21 &  8.5 $\pm$2.1	& 0.06 $\pm$0.34 \\ 
22 &  7.3 $\pm$3.0	& 0.01 $\pm$0.27 \\ 
23 & 11.7 $\pm$10.6 	& 0.24 $\pm$0.25 \\ 
24 &  5.9 $\pm$2.6	& 0.04 $\pm$0.50 \\ 
25 &  3.2 $\pm$2.0	& 0.01 $\pm$0.52 \\ 
26 &  3.8 $\pm$1.1	& 0.02 $\pm$0.20 \\ 
27 &  4.7 $\pm$2.0	& 0.00 $\pm$0.65 \\ 
28 &  5.6 $\pm$1.2	& 0.01 $\pm$0.31 \\ 
\enddata
\end{deluxetable}

\clearpage

\plotone{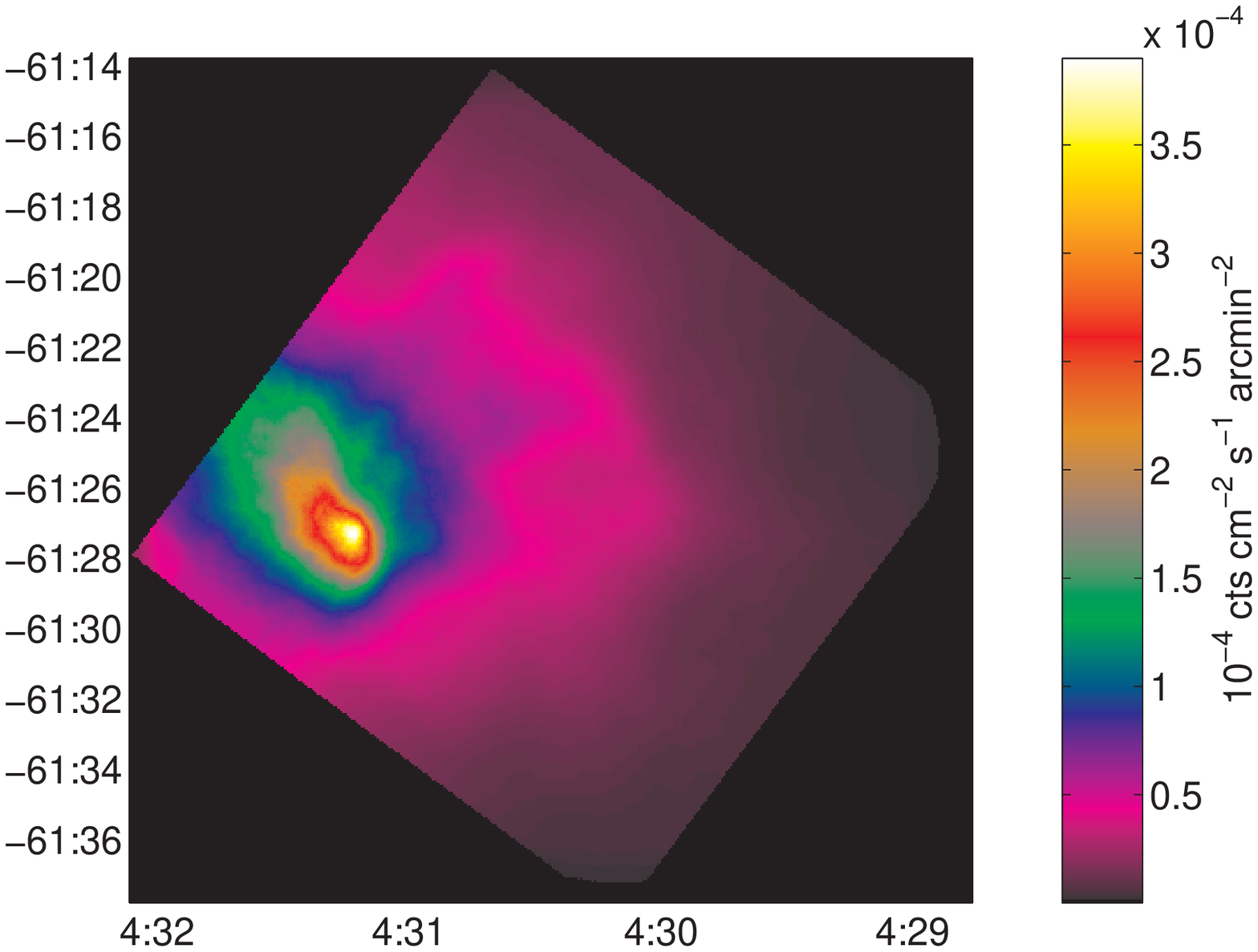}
\figcaption[fig1.eps]{Color coded intensity map of background
subtracted image of the A3266 cluster on the ACIS-I array. Intensity
peak is off center. Elongation is visible toward the NE. Fingers
of emission point to the NW. \label{fig1}}

\plotone{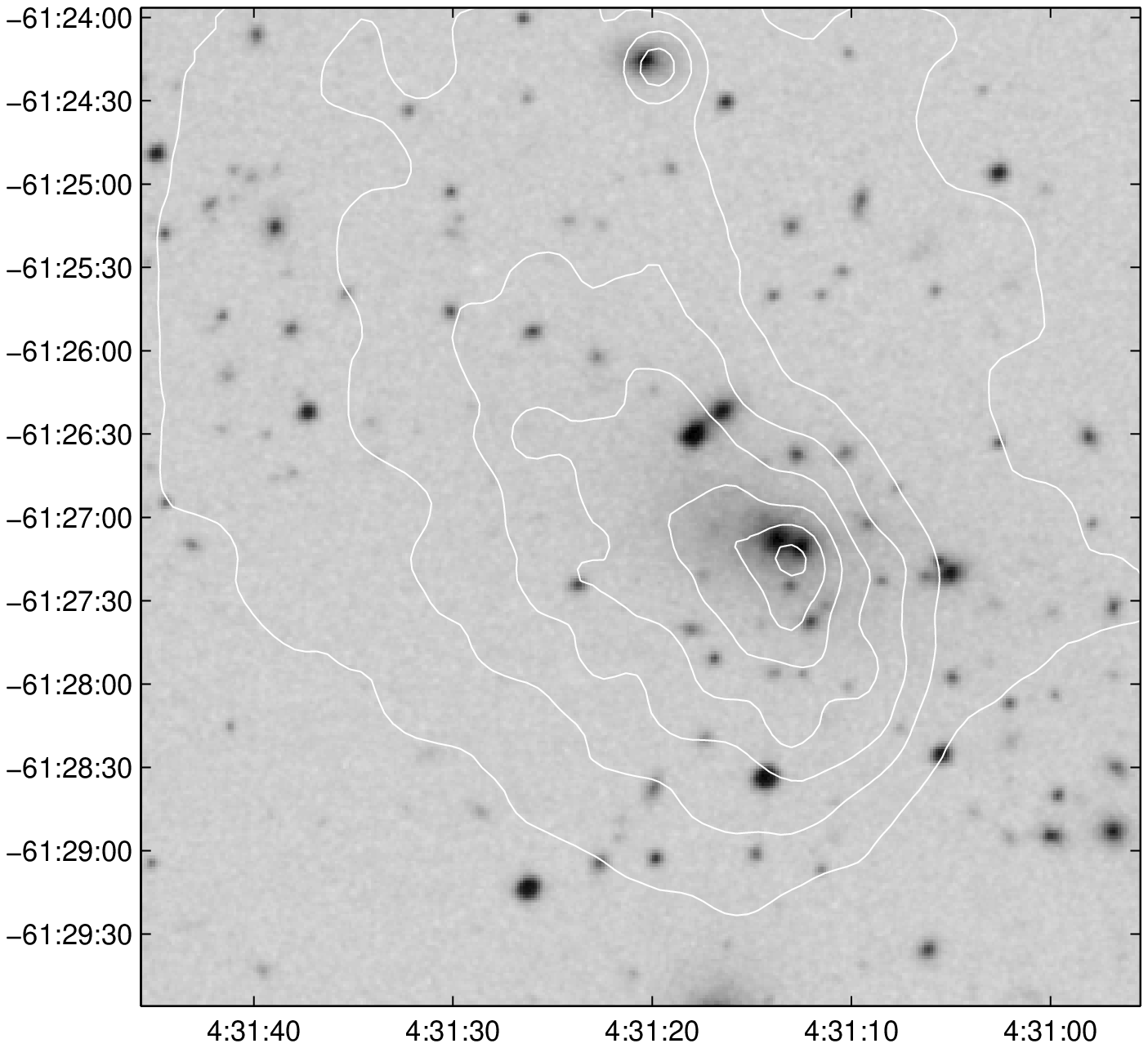}
\figcaption[fig2.eps]{ Contour overlay of the
central cluster emission on chip I1 on the DSS
optical image. Brightness peak is centered on the
central dumbbell  cD galaxy.\label{fig2}}

\plotone{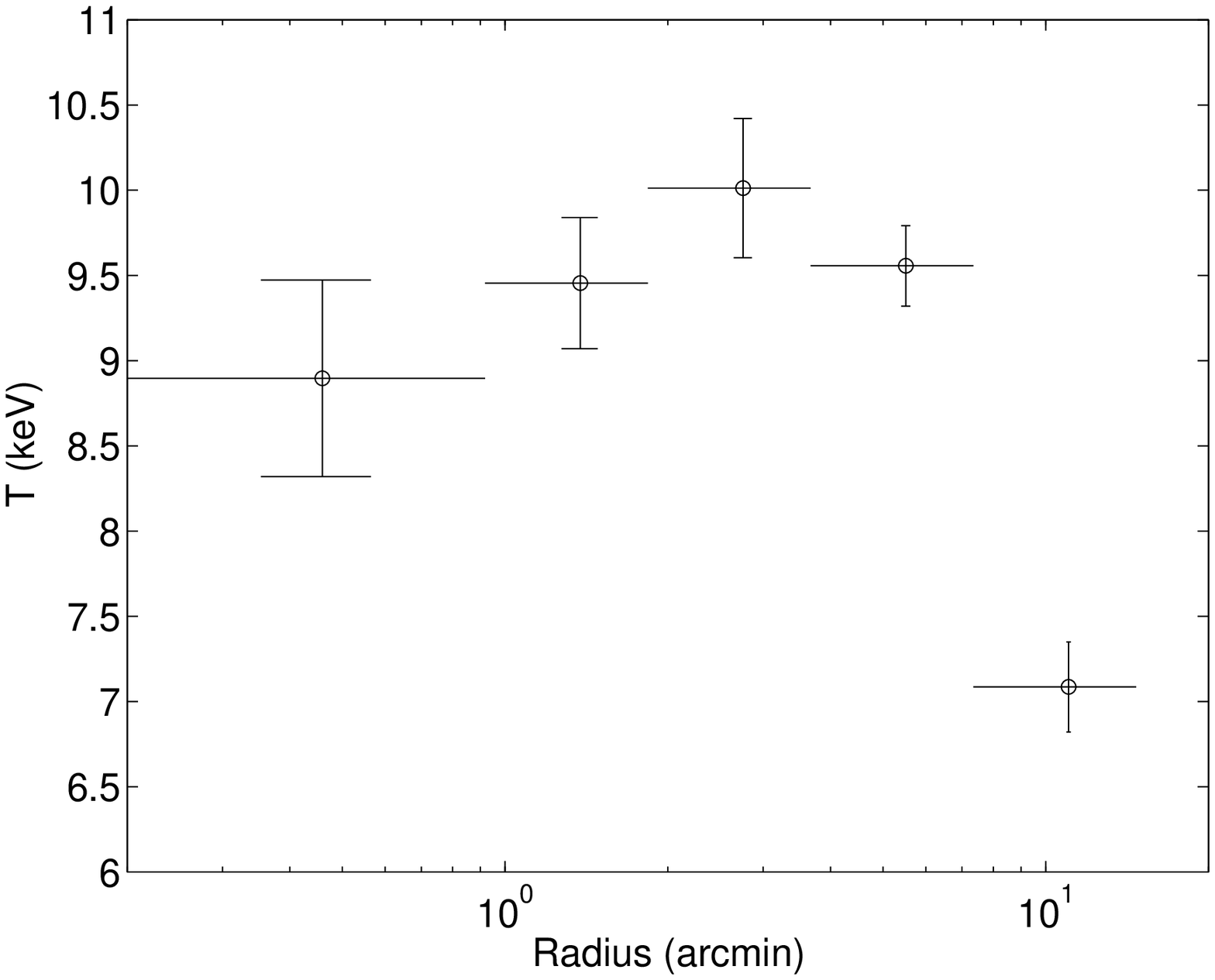}
\figcaption[fig3.eps]{The radial temperature profile shows
a significant temperature drop at larger radius. The central 0 - 1 arcmin
region is likely 2-phase with the cooler cD contribution depressing
the temperature. \label{fig3}}

\plotone{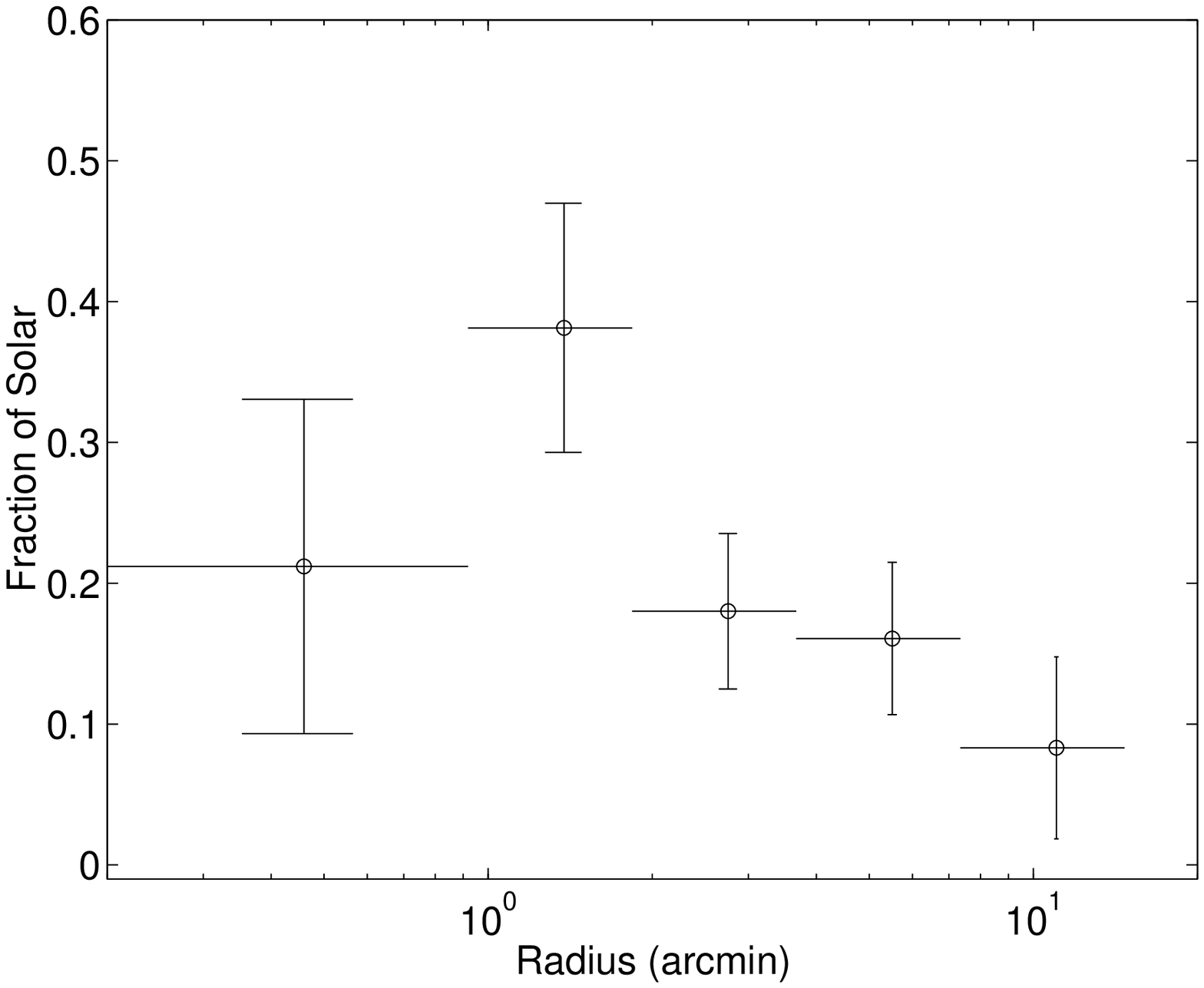}
\figcaption[fig4.eps]{Radial Fe abundance map shows a significant reduction
from the 1 - 2 arcmin annulus and the outer, 7 - 10.5 arcmin annulus.
\label{fig4}}

\plotone{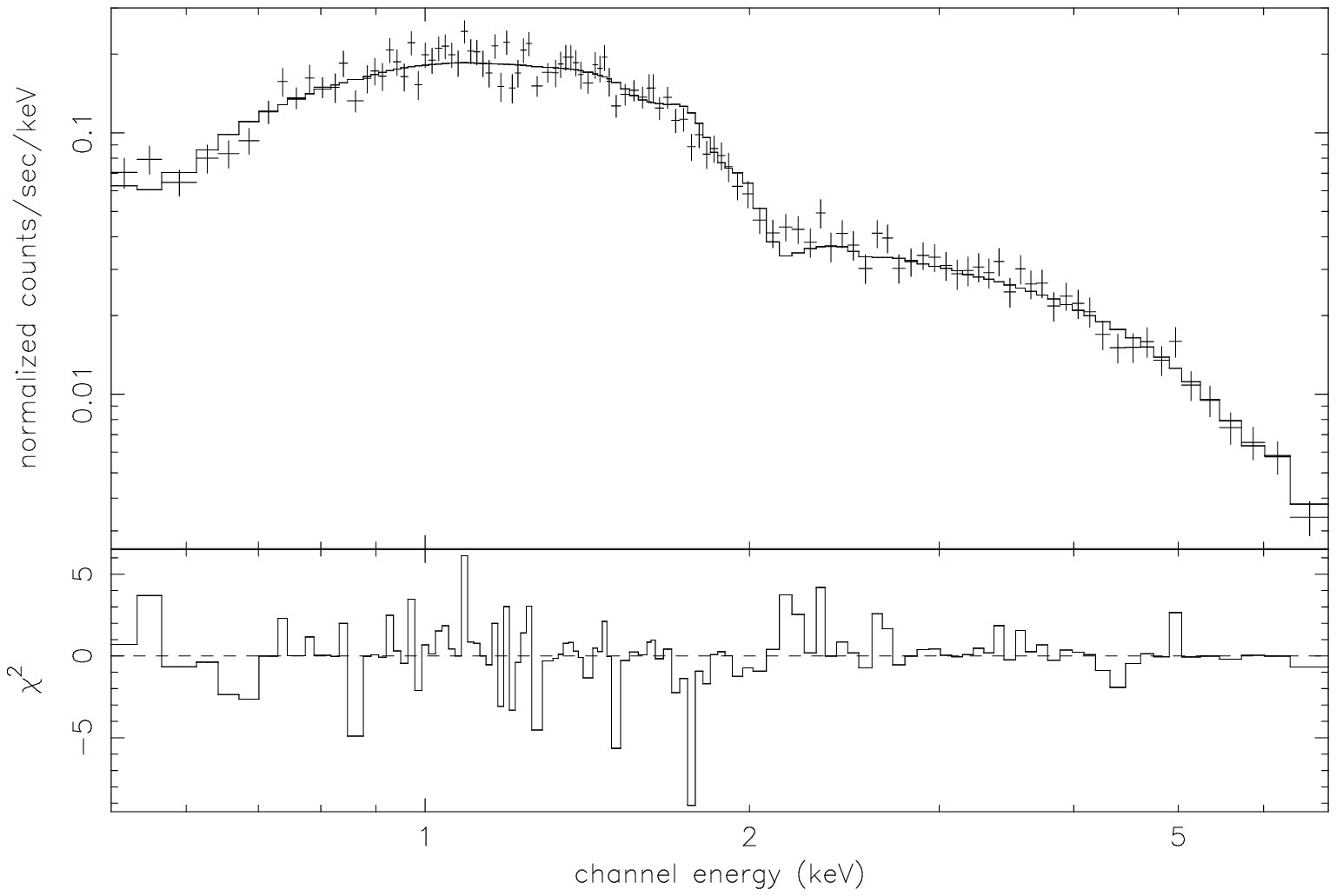}
\figcaption[fig5.eps]{Residuals are shown for a single
thermal component model fit to the spectrum from the inner 0.9
arc min. This region has a lower abundance than the second region.
No residual emission from FeL is visible that would indicate 
a cool, unmodeled gas component.\label{fig5}}

\plotone{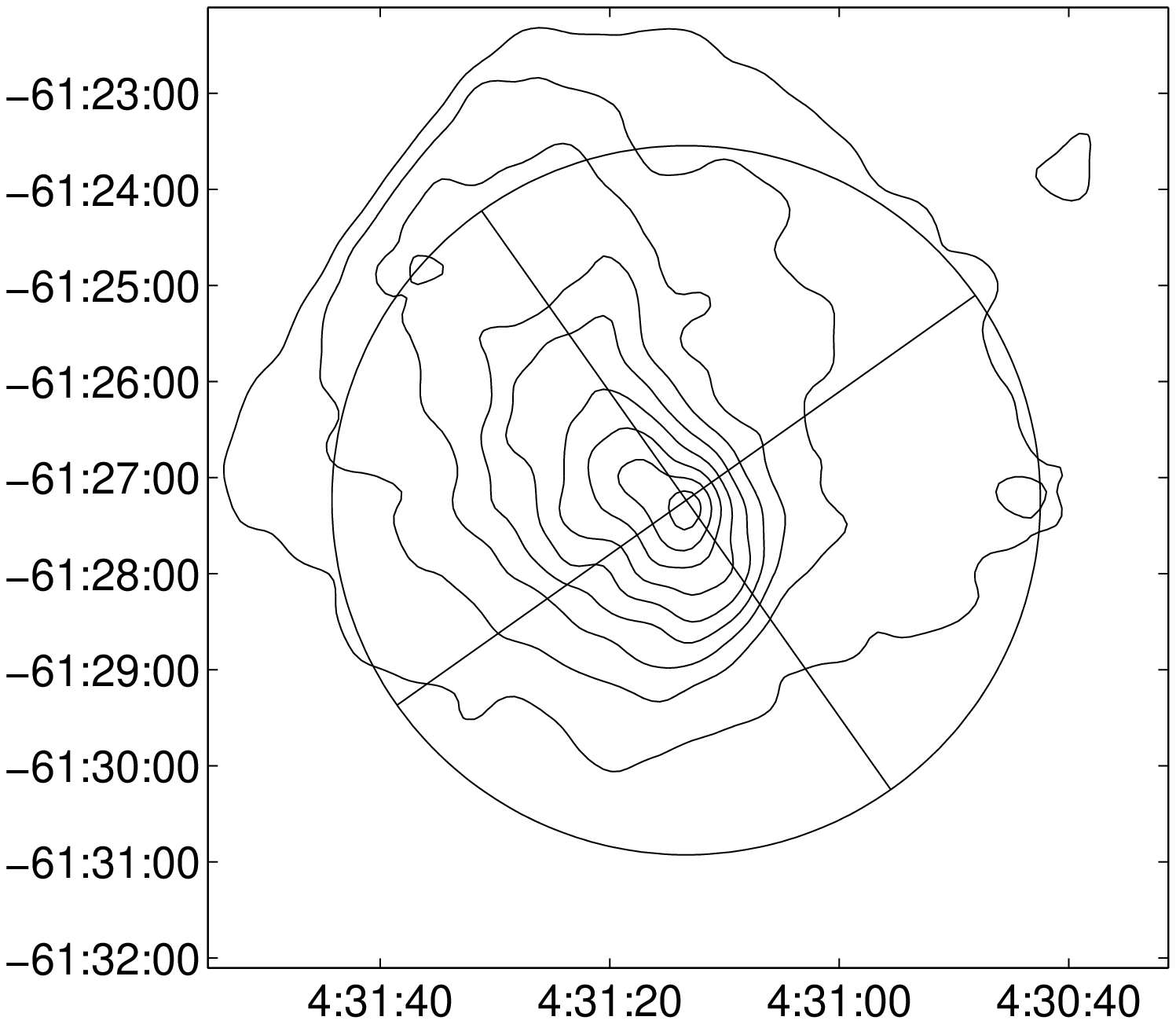}
\figcaption[fig6.ps]{ The temperature and abundance for these
4 quadrants are given in Table 3. The Fe abundance is slightly higher in 3 of the
quandrants and the temperatue slightly lower in the NE. \label{fig6}}

\plotone{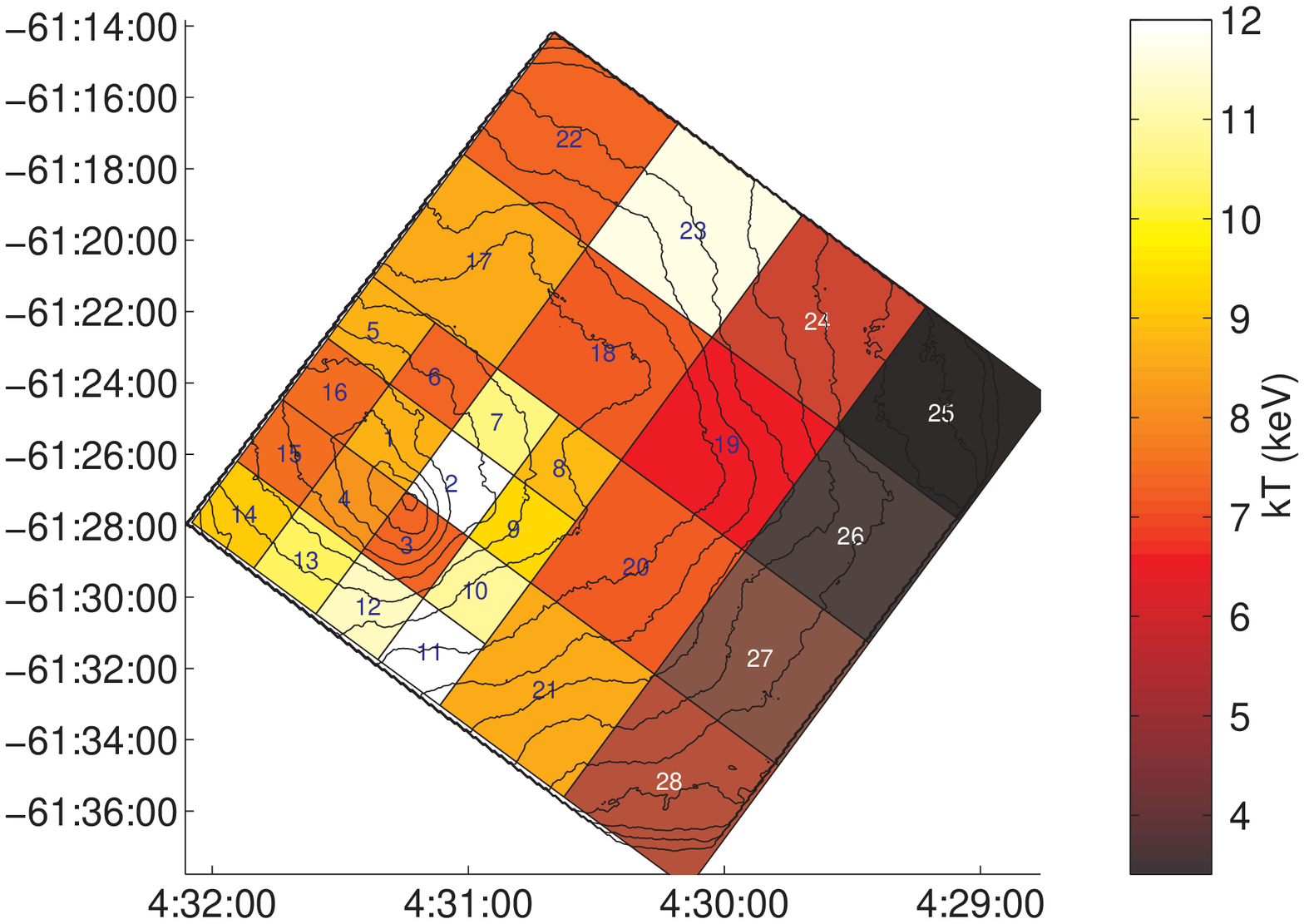}
\figcaption[fig7.eps]{Color coded temperature map with intensity
contours overlayed. The regions identified
correspond to the temperatures given in Table 4. \label{fig7}}

\plotone{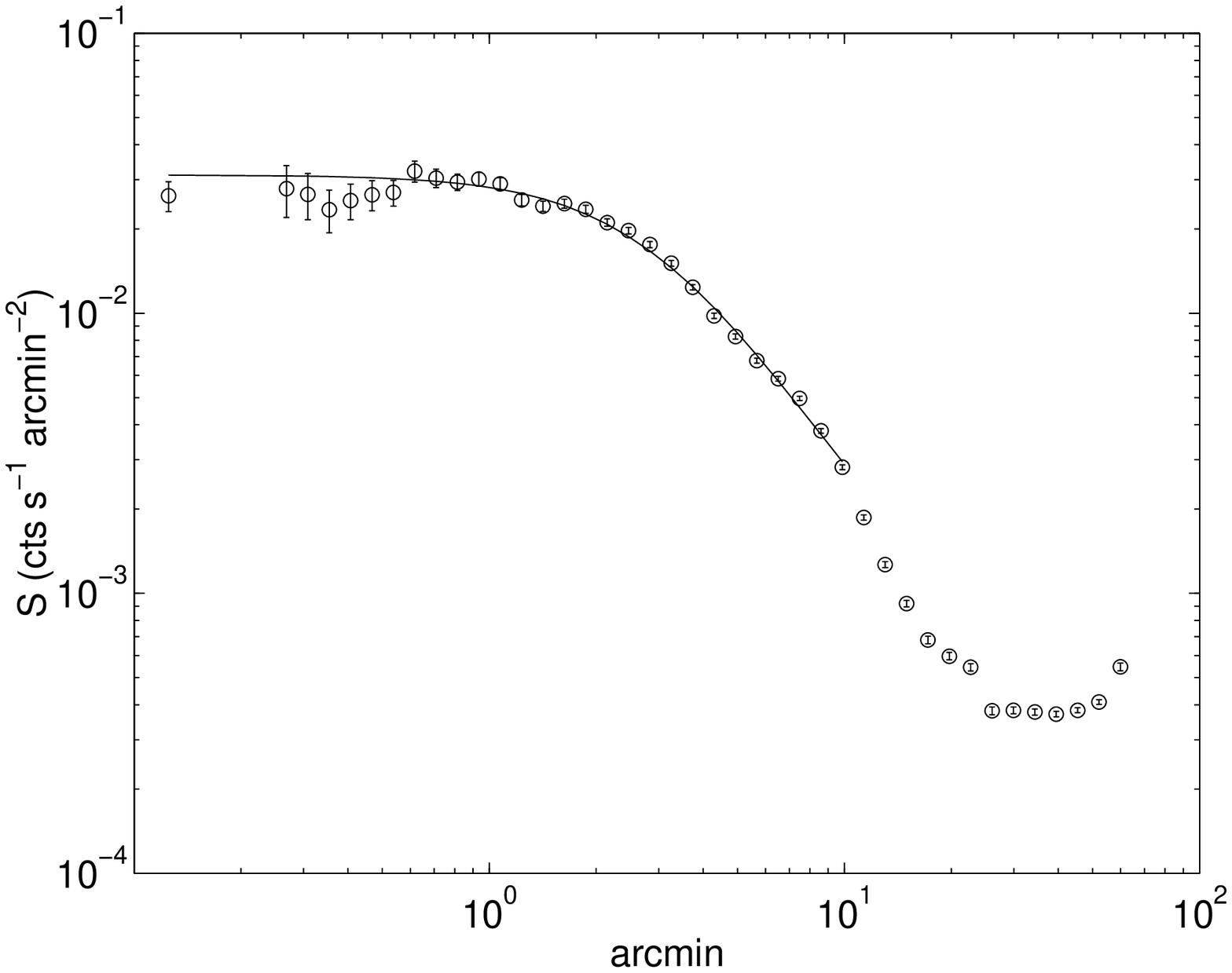}
\figcaption[fig8.eps]{A Beta-model fit to the ROSAT PSPC
radial profile out to 10 arcmin. No significant departures
from a single beta-model are seen at radius less than 10 arcmin.\label{fig8}}

\plotone{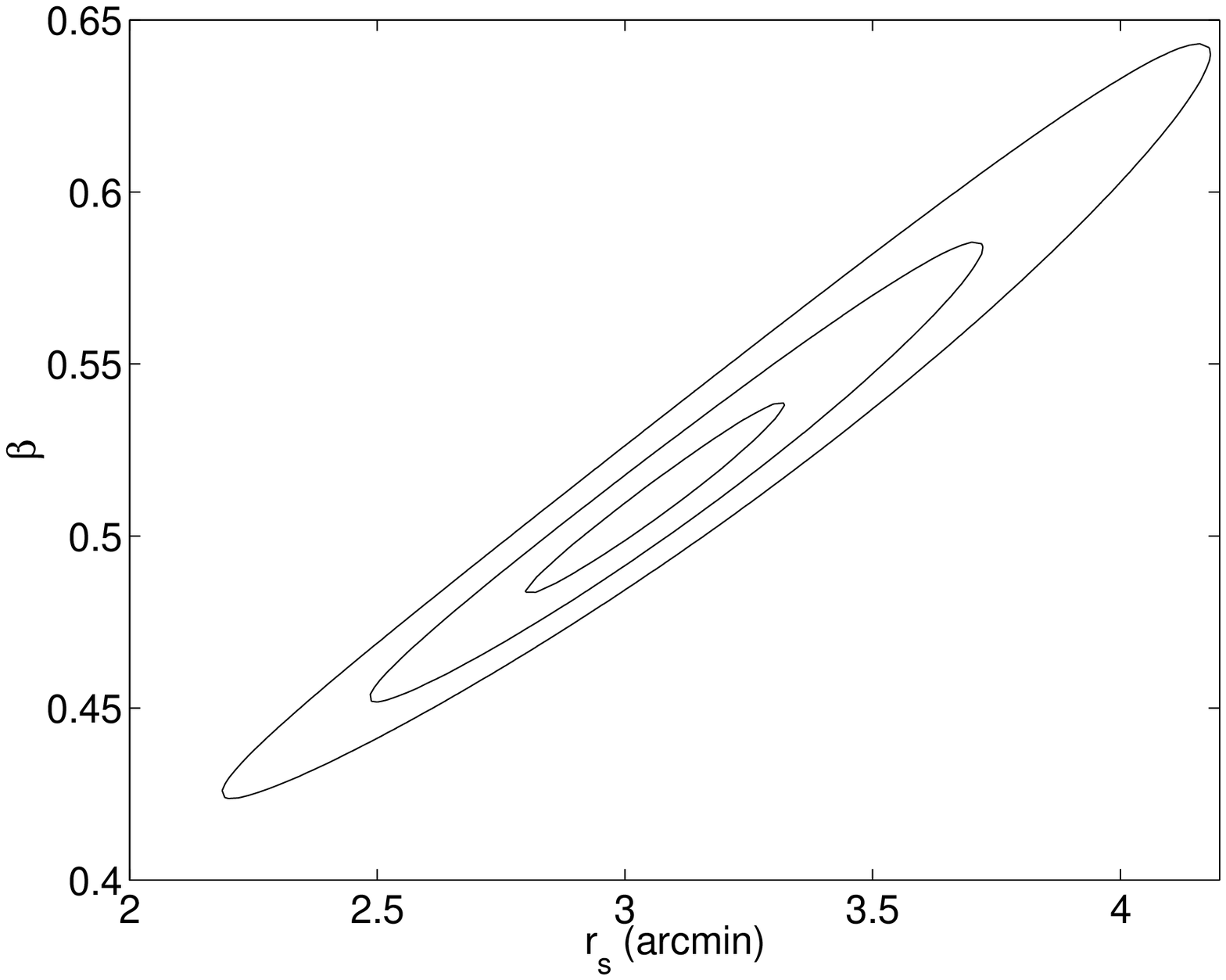}
\figcaption[fig9.eps]{The beta vs. core radius confidence
contours for Abell 3266 indicate a range of paramters that
is typical of clusters within $\sim$1 Mpc.\label{fig9}}

\plotone{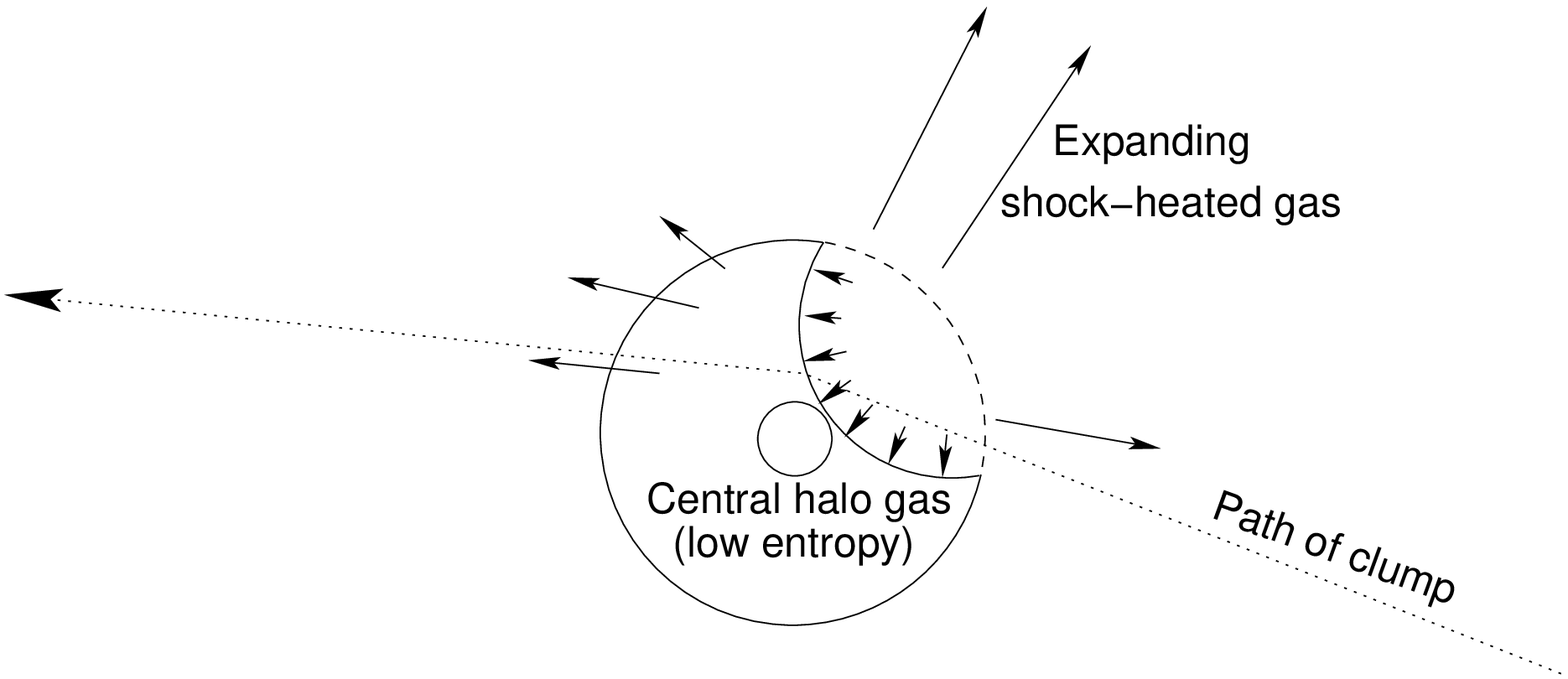}
\figcaption[fig10.eps]{Shockheating of the main cluster ICM
from the merger accompanies the deposition of high abundance subcluster
ICM. \label{fig10}}

\plotone{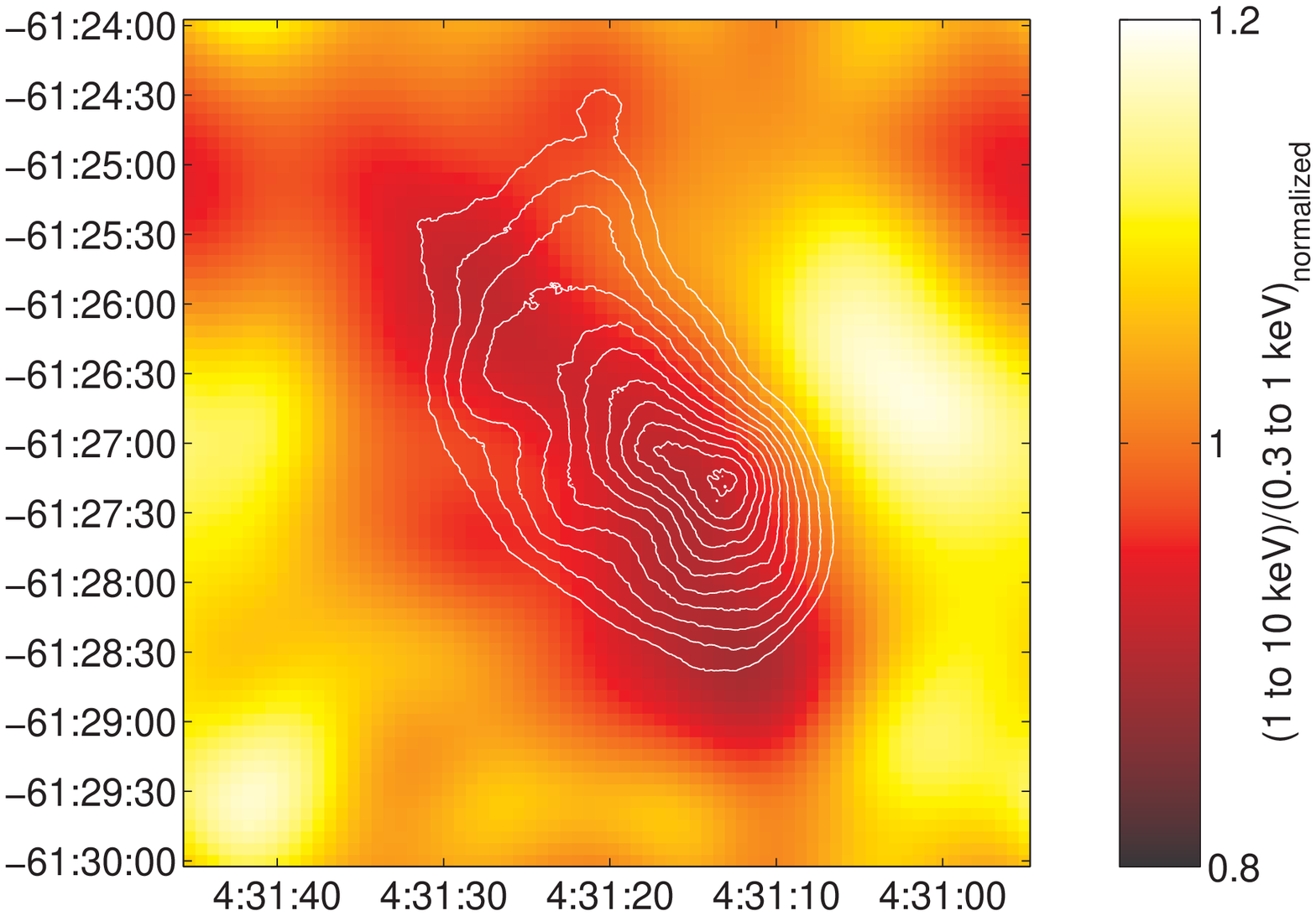}
\figcaption[fig11.eps]{Contour overlayed on hardness map of the 
central region shows a long filament of cool gas running through
the central cD along the merger axis and to the SW. \label{fig11}}

\end{document}